\documentclass[aps,prl,showkeys,twocolumn,groupedaddress]{revtex4-1}
\usepackage{graphicx}
\usepackage{upgreek}
\usepackage{mathtools}
\usepackage{soul}
\usepackage{amsmath}
\usepackage{booktabs}
\usepackage{amssymb}
\usepackage{mathrsfs}
\usepackage{acronym}

\begin{document}


\title{Demonstration of a self-pulsing photonic crystal Fano laser}


\author{Yi Yu, Weiqi Xue, Elizaveta Semenova, Kresten Yvind, and Jesper Mork}
\email[]{jesm@fotonik.dtu.dk, yiyu@fotonik.dtu.dk}
\affiliation{DTU Fotonik, Department of Photonics Engineering, Technical University of Denmark DK-2800, Kongens Lyngby, Denmark}



\begin{abstract}
\noindent Semiconductor lasers in use today rely on mirrors based on the reflection at a cleaved facet or Bragg reflection from a periodic stack of layers. Here, we demonstrate an ultra-small laser with a mirror based on the Fano resonance between a continuum of waveguide modes and the discrete resonance of a nanocavity. The Fano resonance leads to unique laser characteristics. Since the Fano mirror is very narrow-band compared to conventional lasers, the laser is single-mode and in particular, it can be modulated via the mirror. We show, experimentally and theoretically, that nonlinearities in the mirror may even promote the generation of a self-sustained train of pulses at gigahertz frequencies, an effect that was previously only observed in macroscopic lasers. Such a source is of interest for a number of applications within integrated photonics.
\end{abstract}

\maketitle

\noindent Conventional semiconductor lasers mirrors are based on a cleaved facet \cite{SiegmanBOOK} or a Bragg grating, or a two-dimensional
grating resonance \cite{VCSEL,PainterScience,KS,NTT_PHCLaser,PHCJ,WZhouNP,SMB}. In this work, we demonstrate a new concept for lasers, an ultra-small laser with a mirror based on the Fano resonance between a continuum of waveguide modes and the discrete resonance of a nanocavity. The rich physics of Fano resonances \cite{Fano} has recently been explored in a number of different photonic and plasmonic systems \cite{MFano,LFano}. The Fano resonance leads to unique laser characteristics and furthermore represents a very rich dynamical system, which is still to be explored. In particular, since the Fano mirror is very narrow-band compared to conventional lasers, the laser is single-mode and it can be modulated via the mirror. We show, experimentally and theoretically, that nonlinearities in the mirror may even promote the generation of a self-sustained train of pulses at gigahertz frequencies, an effect that was previously only observed in macroscopic lasers \cite{Keller_QS,QS1,OL_QS,JQE_DFB_QS}.

The photonic crystal Fano laser (FL) concept is illustrated in Fig. 1(a). The laser cavity is composed of a line-defect waveguide in a photonic crystal (PhC) membrane and two mirrors. The left mirror is a conventional PhC mirror, realized by blocking the PhC waveguide (WG) with air holes \cite{NODA_Q}. In contrast, the right mirror is due to a Fano interference between the continuum of waveguide modes and the discrete resonance of a side-coupled nanocavity \cite{FANJOSAB}. At resonance, the paths of light through the nanocavity and through the waveguide interfere destructively, leading to a high reflectivity, see Fig. 1(b). If the quality factor ($Q$-factor) of the nanocavity is dominated by its coupling to the waveguide, rather than by intrinsic losses, the maximum reflectivity of the Fano mirror approaches unity, which is the basis for its use as a laser mirror (see Appendix A.1.). This FL concept was suggested in \cite{JMFL} where it was highlighted that, if such a laser could be realized, it should enable modulation not limited by the relaxation oscillations generic to lasers. Here, we report the first experimental demonstration of such a Fano laser, including its unique mode-selection properties compared to usual line-defect PhC lasers. Furthermore, we report the discovery that the laser can operate in a regime, where nonlinearities in the nanocavity induce a self-sustained train of optical pulses, even in the absence of external modulation. The experimental results are well explained by a theoretical model that takes into account the characteristics of the Fano mirror, which leads to rich dynamical behavior.
\begin{figure*}[!htb]
\centering
\includegraphics[width=5.5in]{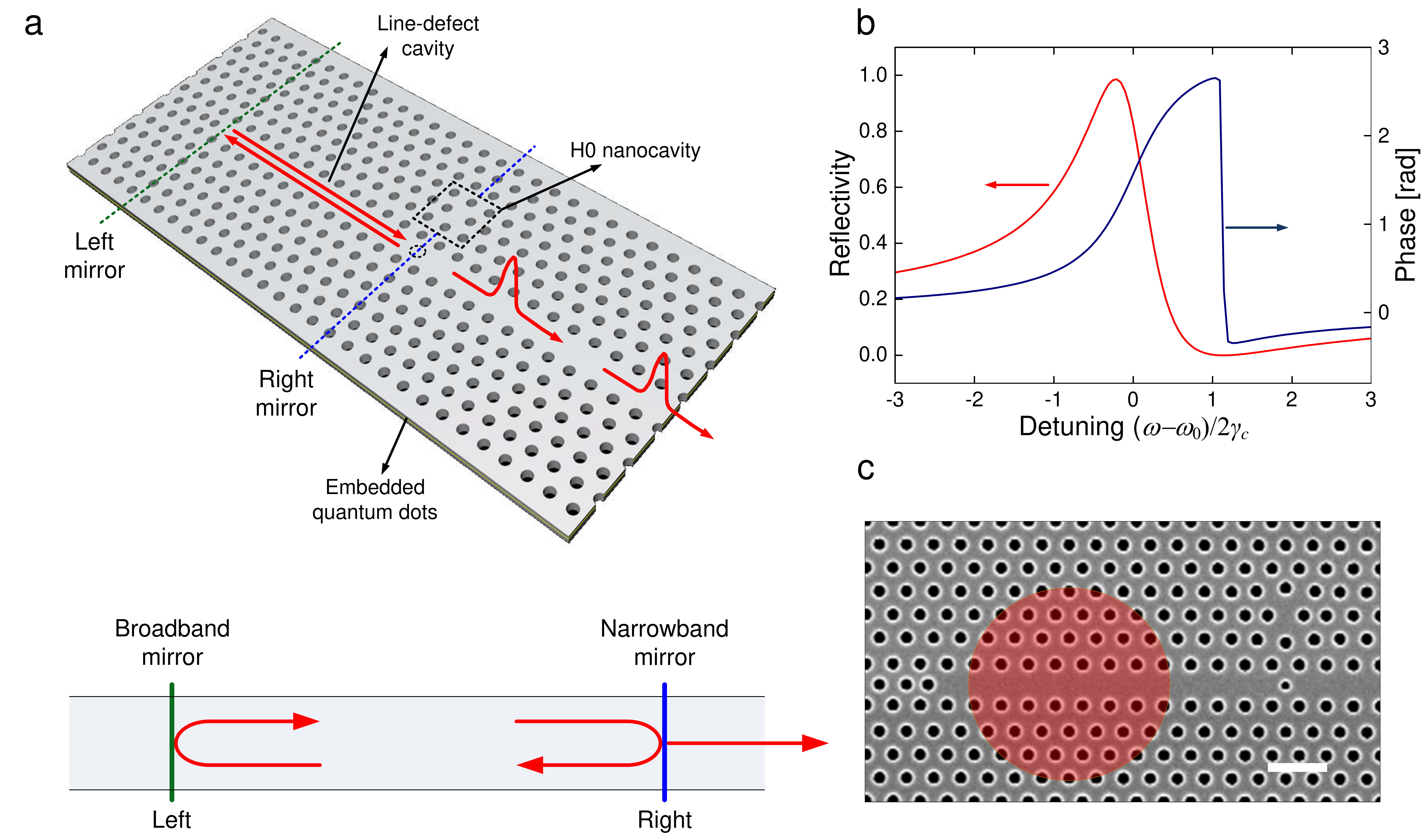}%
\caption{\label{fig:F1} (a). Schematic of Fano laser based on an InP photonic-crystal membrane which contains an active region with 3 layers of InAs quantum dots. The laser cavity is composed of a photonic-crystal line-defect waveguide, terminated by a broadband left mirror and a narrowband right mirror, realized by Fano interference between a waveguide continuum mode and a discrete mode in a side-coupled H0 nanocavity. (b). Calculated reflectivity (red line) and phase (blue line) of the nanocavity mirror as seen from the waveguide with $r_{B}$=0.4. $\omega_{0}$ and 2$\gamma_{c}$ are the frequency and bandwidth of the nanocavity resonance, respectively. The coupling and intrinsic quality-factor of the nanocavity is 800 and 100000, respectively. (c). SEM image of a fabricated sample. The scale bar corresponds to 1 $\upmu $m and the red disk indicates the optical pumping region.}
\end{figure*}

A related passive structure was used for demonstrating dynamic tuning of cavity $Q$-factor \cite{NODAM}, but in our case, an active material composed of 3 layers of InAs quantum dots (QDs) is incorporated inside the membrane and the structures are designed such that positive net material gain can be achieved by optical pumping (see Appendix B.1. for details of the fabricated structure). In steady state, the lasing mode at frequency $\omega$ fulfills the oscillation condition (see Appendix A.1. for the detailed FL model)
\begin{equation}\label{E1}
{{r}_{L}}{{r}_{R}}\exp[2ikL]=1\
\end{equation}
where $r_{L}$ and $r_{R}$ are the left and right mirror field reflection coefficients, respectively, $k$ is the complex wavenumber of the L-cavity, and $L$ is the effective length of the cavity.
Eq. (1) leads to conditions for the gain and wavelength of oscillating modes. In the present case of a very dispersive mirror, these equations are strongly coupled. Thus, the conditions of minimum threshold gain and the phase being an integer multiple of 2$\pi$ are only fulfilled simultaneously for specific cavity lengths. This, in turn, means that the laser can be efficiently modulated via the mirror, e.g. by changing its resonance frequency or peak reflectivity. Here, we employ a nanocavity of the H0-type, which has the smallest mode volume of all known PhC defect cavities \cite{MinQiu}. Another fundamental control of the laser is provided via the shape parameter $q$ of the Fano resonance \cite{MFano}, which can be controlled via an airhole with variable radius added below the nanocavity \cite{MHOL,YYLPR} (see the black dashed circle in Fig. 1(a)). The airhole thus serves as a partially transmitting element with a variable reflection coefficient $r_{B}$. Thus, $r_{B}$=0 ($r_{B} \neq$0) corresponds to a symmetric (asymmetric) Fano resonance with $q$=0 ($q \neq$0).
\begin{figure}
\centering
\includegraphics[width=2.5in]{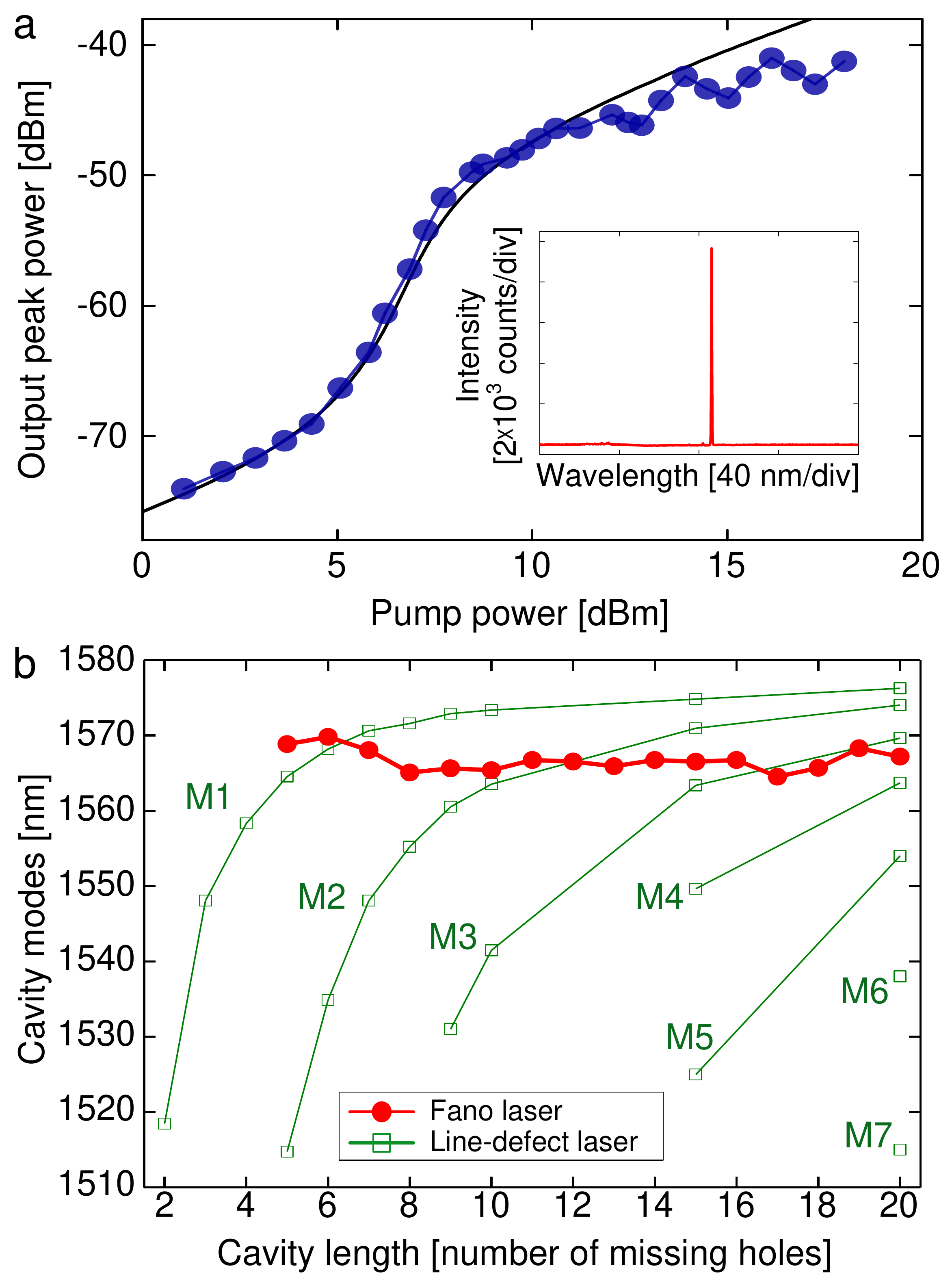}%
\caption{\label{fig:F2} (a). Measured output peak power versus pump power. The blue dotted line is the experimental data and the black solid line is the theoretical fit. Inset: Optical spectrum for a pump power of $\sim$7.5 dBm. (b). Measured cavity mode wavelengths versus cavity length for Fano lasers (red markers) and ordinary photonic-crystal line-defect cavity lasers (green markers). M1-M7 denote the mode order, M1 being the fundamental mode.}
\end{figure}
In the following investigation, we focus on a FL with $r_{B}^{2}\approx 0.16$ and a cavity length corresponding to 12 missing holes between the left mirror and the nanocavity. Using vertical pumping with a continuous-wave (CW) light source at 1480 nm and vertical collection (see Appendix B.2.), we first characterize the static properties of the FL. Only the waveguide region is pumped, cf. Fig. 1(c) and due to the quantum confinement of QDs, the generated free carriers are well confined. The measured maximum output power is of the order -40 dBm due to the detection of vertically scattered light only. This can be improved considerably by employing an efficient coupler to couple the light out in the in-plane direction, as appropriate for on-chip applications and already demonstrated using conventional PhCs \cite{PhC_Coupler}. As seen in the inset of Fig. 2(a), a single dominant mode at 1566 nm is observed and the corresponding peak power (Fig. 2(a)) vs pump power, shows a clear transition to lasing at a threshold pump power of 4.6 mW. The mode properties of the laser are investigated by measuring the variation of the lasing wavelength for a number of lasers with different cavity lengths, cf. Fig. 2(b). For comparison, we have also shown results for conventional PhC line defect lasers with both left and right mirror being composed of a terminated waveguide \cite{Arakawa}. The investigated FLs were found to be single-mode with a wavelength nearly independent of the cavity length. This is in stark contrast to the ordinary PhC L-cavity lasers where several longitudinal modes are observed \cite{PRL_WQ}, and all the modes shift to longer wavelengths for increasing cavity length consistent with the modes approaching the PhC Brillion-zone-edge. This demonstrates the good mode selection properties of FLs enabled by the narrowband mirror. Furthermore, the pinning of the laser wavelength by the nanocavity resonance avoids the transition into the slow-light region with increasing laser length, which leads to an unconventional increase of laser threshold with cavity length \cite{PRL_WQ}.

Considering the strict requirement of fulfilling the phase matching condition close to the mirror reflectivity peak, it appears surprising that all the investigated FLs, representing a wide range of cavity lengths, actually worked. In addition to a careful design of the PhC hole pattern, we believe that the ability to slightly change the refractive index profile via the pumping spot position is important in this respect. And, as shown in Fig. 1(b), in the vicinity of the mirror reflection peak, the associated large phase variation helps to fulfill the phase matching condition.

\begin{figure}
\centering
\includegraphics[width=3.5in]{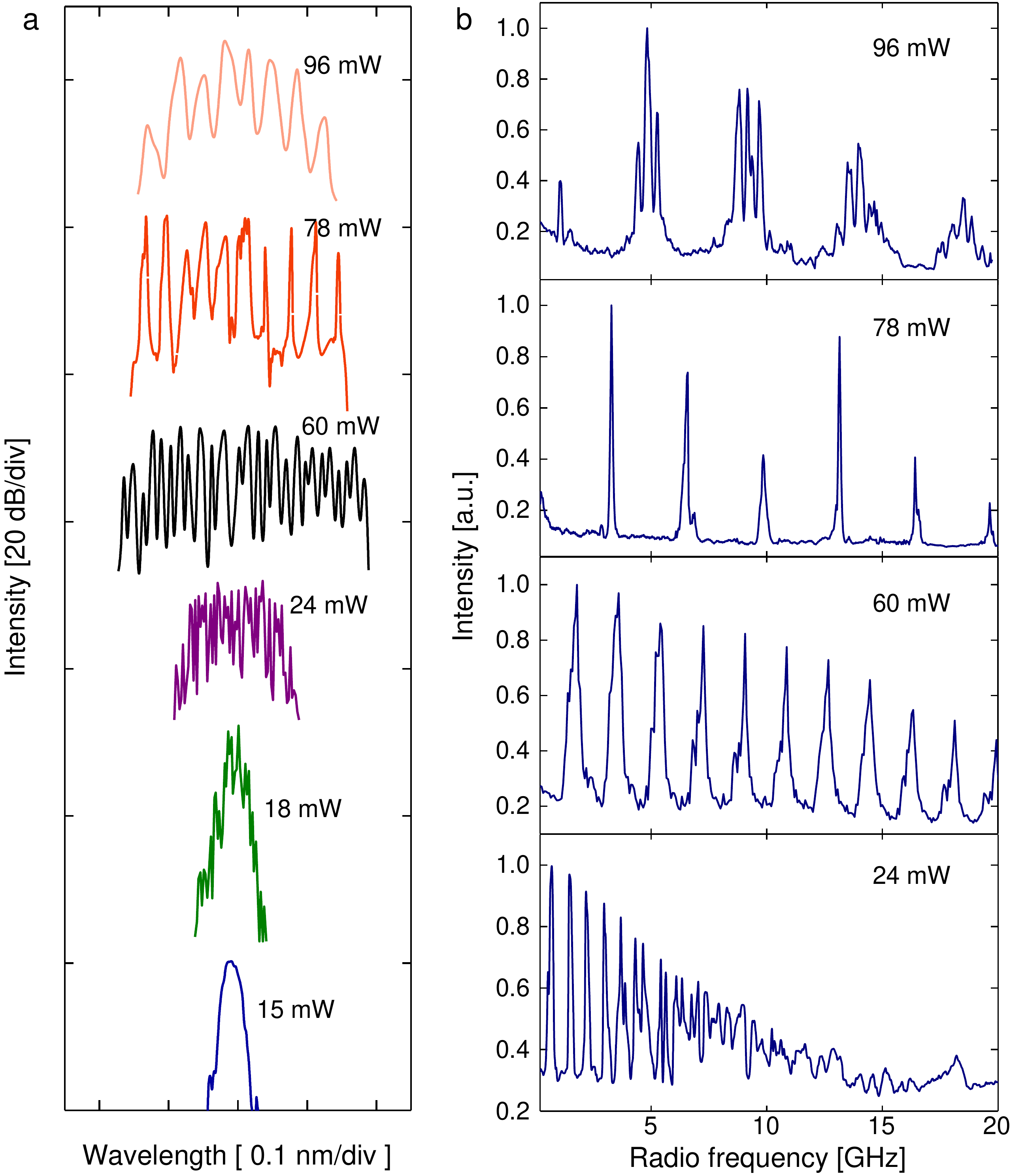}%
\caption{\label{fig:F3} (a). Measured optical spectra of the Fano laser for pump powers of 15, 18, 24, 60, 78 and 96 mW, respectively. For clarity, the spectra are vertically shifted. (b). Measured radio frequency spectra of the laser power for pump powers of 24, 60, 78 and 96 mW, respectively.}
\end{figure}

The evolution of the optical and radio-frequency (RF) spectra of the laser intensity is recorded for increasing pump power levels, see Fig. 3. The optical spectrum broadens with pump power, accompanied by the appearance of multiple sidebands, and at the same time, the RF spectrum displays a comb of equidistant peaks. These observations strongly suggest that the laser undergoes a transition from CW lasing to the generation of a self-sustained train of short optical pulses. This interpretation is strongly supported by dynamical simulations of the developed FL model (see Appendix A.2.). An example simulation is shown in Fig. 4(a). The onset of pulsing is initiated by a transient laser intensity spike, which leads to carrier generation in the nanocavity via linear absorption in the unpumped quantum dots. These carriers subsequently saturate the loss and increase the Fano mirror reflectivity, which provides positive feedback to the intensity fluctuation. Eventually a dynamical equilibrium state of short pulses is formed (see Appendix A.2.). As far as we are aware, this is the first observation of passive pulse generation in an ultra-compact PhC laser. The pulse-generation mechanism bears resemblance to the use of semiconductor-saturable mirrors for passive $Q$-switching of external cavity lasers \cite{Keller_QS} and multi-section DFB lasers \cite{OL_QS,QS1,JQE_DFB_QS}, but besides the extreme miniaturization enabled by our configuration; the Fano resonance adds additional richness and design opportunities.
\begin{figure*}
\centering
\includegraphics[width=6.0in]{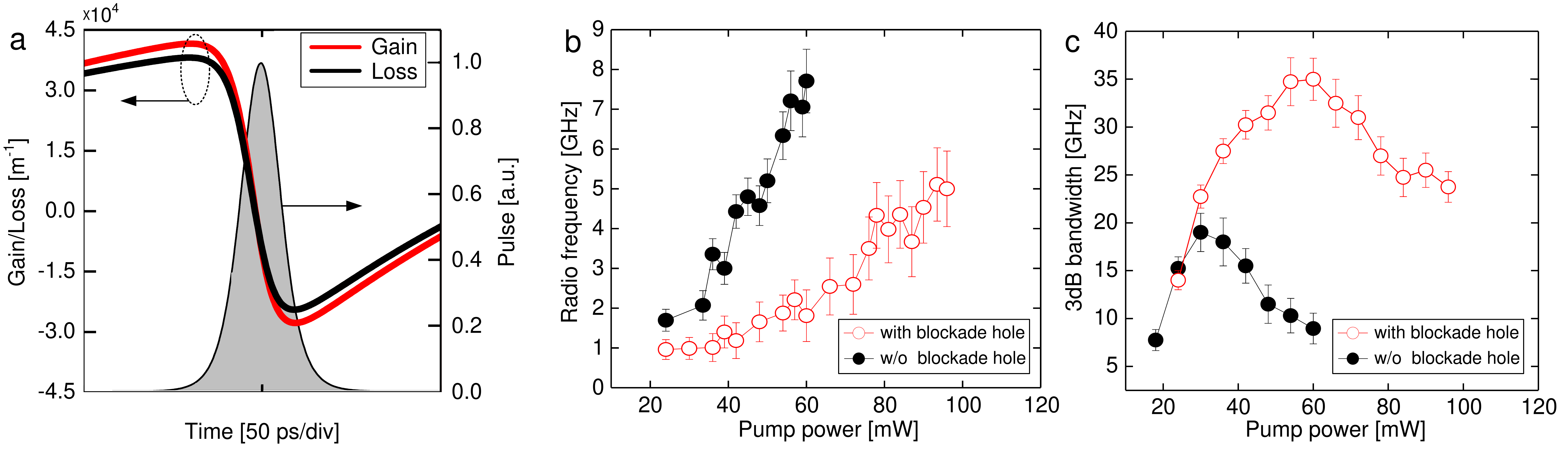}%
\caption{\label{fig:F4} (a). Simulated time evolution of the gain (red curve), and loss (black curve) changes and laser output (gray shade) of the Fano laser when self-pulsations sets in. (b). Pulsation frequency and (c). 3 dB bandwidth of optical spectra versus pump power. The error bars are the standard deviation of the measurements. Red (black) symbols correspond to a Fano laser with (without) a blockade hole in the waveguide.}
\end{figure*}

Fig. 4(b) shows the measured pulse repetition frequency versus pump power for FLs with and without a blockade hole. For both types of FL, self-pulsation sets in at a pump power of $\sim$20 mW, and the repetition frequency increases with pump power. The measured highest repetition frequency is 8 GHz, obtained for the laser without a blockade hole. The self-pulsation in the laser without a blockade hole is quenched when the pump power exceeds 60 mW, giving a smaller power range than for the laser with a blockade hole. Fig. 4(c) shows the corresponding pump-power variation of the full width at half maximum (FWHM) of the optical spectra. Both laser structures show the same tendency of an initial increase of the FWHM followed by a decrease. The variations of pulse repetition frequency as well as FWHM with pump power are in good qualitative agreement with simulations (see Appendix A.2.). The simulations show that the inverse of the spectral FWHM is indicative of the temporal pulse width, with a pulse width of $\sim$15 ps and a peak power of $\sim$1.2 mW (propagating in the in-plane direction) being achieved for the laser containing a blockade hole, cf. Fig. 4(a).

We have experimentally demonstrated a new concept for nanolasers, where one of the mirrors is provided by a Fano resonance between a waveguide continuum and a discrete resonance in a nanocavity. This type of mirror is narrow-band and highly dispersive, providing new opportunities for modulating the laser via the mirror rather than waveguide cavity. In particular, we showed that the laser is single-mode and the mirror can act as a saturable absorber, leading to pulse-generation. As far as we are aware this is the first demonstration of passive pulse generation in nanolasers. Numerous applications of such FLs may be envisioned. Besides a pulse source for on-chip communications and sampling, the FL may be optically triggered, e.g., for application in all-optical synchronization and switching. Furhermore, it was recently proposed that by modulating a laser via a mirror resonance \cite{JMFL}, it should be possible to realize terahertz frequency modulation bandwidths, not limited by the intrinsic relaxation oscillations limiting conventional modulation schemes.


\section{Acknowledgment}
\noindent We thank L. Ottaviano for assistance in wafer preparation, and H. Hu and L. K. Oxenl\o we for assistance with experimental set-ups. The authors acknowledge financial support from Villum Fonden via the NATEC (NAnophotonics for TErabit Communications) Centre and YIP QUEENs.

\section{Appendix}
\section{A.	Theoretical model}
\noindent  \textbf{A.1. Mirror property}
\begin{figure}[!htb]
\centering
\includegraphics[width=3.4in]{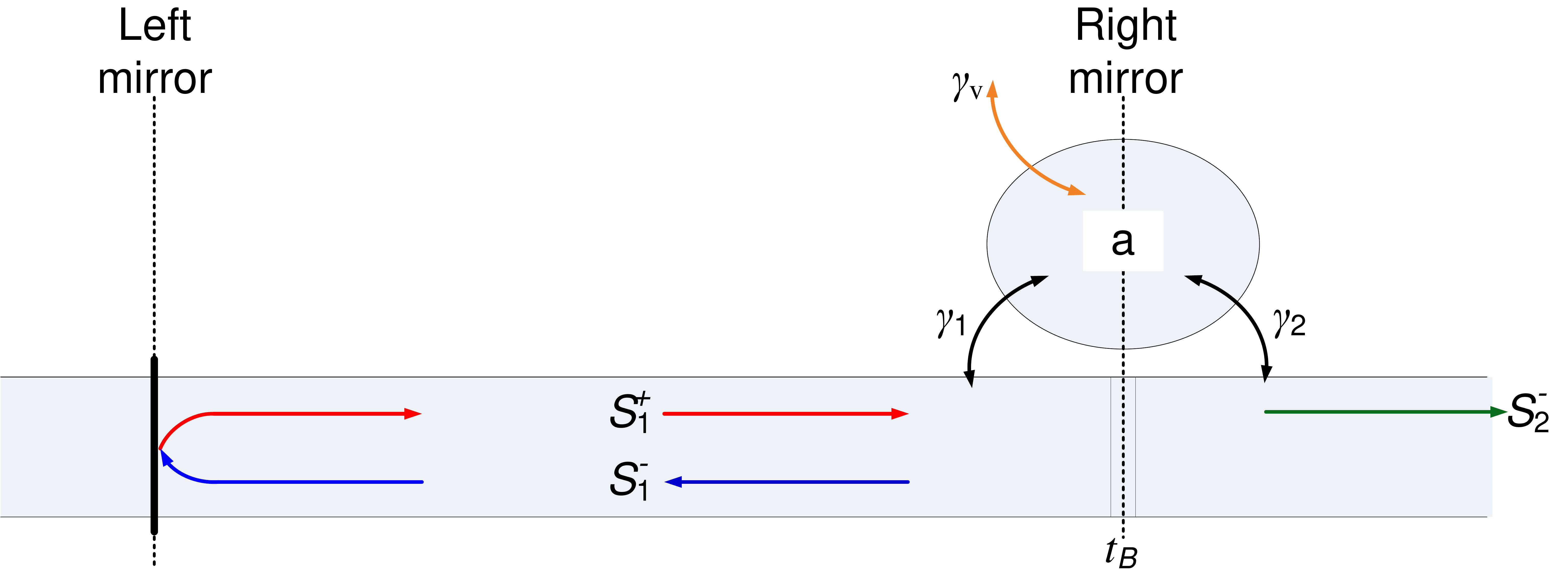}%
\caption{\label{fig:SF1} Schematic of the Fano laser.}
\end{figure}

\noindent The Fano laser (FL) structure (Fig.\ref{fig:SF1}) can be viewed as a line-defect cavity (L-cavity) formed by a line-defect waveguide (WG) in-between two mirrors. The left mirror is broadband while the right mirror is narrowband, realized by Fano interference between a continuum mode in the WG and a discrete mode in a side-coupled nanocavity. A partially transmitting element (PTE) with a transmittance (reflectivity) of ${{t}_{B}}$(${{r}_{B}}$) is added below the nanocavity in the WG. We assume $t_{B}^{2}+r_{B}^{2}=1$ and neglect the out-of-plane scattering of the PTE since it is found negligible for our structures (2 orders of magnitude smaller than the transmittance/reflectivity as estimated by finite-difference time-domain method (FDTD) calculations). The out-of-plane scattering may become important when operating in the slow light region, which is not the case for the structures investigated here, but is an interesting topic for future investigations. The excitation of the nanocavity via the waveguide can be described by temporal coupled mode equations \cite{HausBook} as:

\begin{equation}\label{E1}
\renewcommand\theequation{s\arabic{equation}}
\frac{da(t)}{dt}=(-i{{\delta }_{c}}-{{\gamma }_{t}})a(t)+\sqrt{2{{\gamma }_{1}}}{{e}^{i{{\theta }_{1}}}}s_{1}^{+}(t)\
\end{equation}
\begin{equation}\label{E2}
\renewcommand\theequation{s\arabic{equation}}
s_{1}^{-}(t)={{r}_{B}}s_{1}^{+}(t)+\sqrt{2{{\gamma }_{1}}}{{e}^{i{{\theta }_{1}}}}a(t)\
\end{equation}
\begin{equation}\label{E3}
\renewcommand\theequation{s\arabic{equation}}
s_{2}^{-}(t)=-i{{t}_{B}}s_{1}^{+}(t)+\sqrt{2{{\gamma }_{2}}}{{e}^{i{{\theta }_{2}}}}a(t)\
\end{equation}
Here, ${{\left| a\left( t \right) \right|}^{2}}$ is the energy in the nanocavity, ${{\gamma }_{c}}={{\gamma }_{1}}+{{\gamma }_{2}}$ is the field decay rate due to the coupling between the nanocavity and WG; ${{\gamma }_{t}}={{\gamma }_{v}}+{{\gamma }_{\alpha }}+{{\gamma }_{1}}+{{\gamma }_{2}}$ is the total decay rate of the nanocavity, with ${{\gamma }_{v}}$ accounting for the vertical emission and ${{\gamma }_{\alpha }}$ accounting for the remaining losses, mainly the linear absorption at transparency in the QDs. They are related to the cavity $Q$-factors as ${{\gamma }_{v}}={{\omega }_{0}}/2{{Q}_{v}}$, ${{\gamma }_{\alpha }}={{\omega }_{0}}/2{{Q}_{\alpha }}$ and ${{\gamma }_{c}}={{\omega }_{0}}/2{{Q}_{c}}$ with ${{Q}_{v}}$, ${{Q}_{\alpha }}$ and ${{Q}_{c}}$ being the vertical, absorption and coupling $Q$-factors of the nanocavity, respectively. $s_{1}^{+}(t)$ is the amplitude of the incoming waves (from the left), $s_{1}^{-}(t)$ and $s_{2}^{-}(t)$ are the amplitudes of the reflected and output waves, with ${{\left| s_{1,2}^{\pm }(t) \right|}^{2}}$ being the corresponding powers. In Eq. (s\ref{E1}), ${{\delta }_{c}}={{\omega }_{0}}+\Delta {{\omega }_{c}}-\omega $ is the frequency detuning between the nanocavity resonance frequency, ${{\omega }_{0}}+\Delta {{\omega }_{c}}$, and the frequency of the right-propagating signal $\omega $, where ${{\omega }_{0}}$ is the unperturbed resonance frequency. Furthermore, $\Delta {{\omega }_{c}}$ is the complex time-dependent change of the nanocavity resonance
\begin{equation}\label{E4}
\renewcommand\theequation{s\arabic{equation}}
\Delta {{\omega }_{c}}={{K}_{D}}\cdot {{N}_{c}}(t)-i{{K}_{A}}\cdot {{N}_{c}}(t)\
\end{equation}
The real part of $\Delta {{\omega }_{c}}$ accounts for the resonance frequency shift, with ${{K}_{D}}$ being the nonlinear coefficient relating to a combination of free carrier dispersion and bandfilling. The imaginary part of $\Delta {{\omega }_{c}}$ accounts for absorption in the nanocavity, with the coefficient ${{K}_{A}}$ accounting mainly for linear absorption. ${{N}_{c}}(t)$ is the mode averaged carrier density in the nanocavity (mainly in the QDs) whose dynamics can be approximately described as
\begin{equation}\label{E5}
\renewcommand\theequation{s\arabic{equation}}
\frac{d{{N}_{c}}(t)}{dt}=-R\left( {{N}_{c}}(t) \right)+{{G}_{LA}}\left( {{N}_{tr}}-{{N}_{c}}(t) \right){{\left| a(t) \right|}^{2}}\
\end{equation}
In Eq. (s\ref{E5}), ${{G}_{LA}}$ is the carrier generation coefficient due to linear absorption. ${{N}_{tr}}$ is the carrier density at transparency. $R\left( {{N}_{c}}(t) \right)$ is the carrier recombination rate and for simplicity, we assume $R\left( {{N}_{c}}(t) \right)\simeq {{N}_{c}}(t)/{{\tau }_{c}}$ with ${{\tau }_{c}}$ being the effective carrier lifetime.
At steady state, from Eq. (s\ref{E1}), the reflection coefficient of the Fano mirror can be obtained as
\begin{equation}\label{E6}
\renewcommand\theequation{s\arabic{equation}}
{{r}_{R}}\left( \omega  \right)={{r}_{B}}+\frac{2{{\gamma }_{1}}{{e}^{2i{{\theta }_{1}}}}}{i{{\delta }_{c}}+{{\gamma }_{t}}}\
\end{equation}
By exploiting energy conservation and time-reversal symmetry \cite{HausBook}, we get the relations between the phase of the coupling coefficient and the other measurable quantities as ${{e}^{i{{\theta }_{1}}-i{{\theta }_{2}}}}=\frac{\sqrt{{{\gamma }_{1}}}}{\sqrt{{{\gamma }_{2}}}}\frac{1}{i{{t}_{B}}}\left( {{e}^{2i{{\theta }_{1}}}}+{{r}_{B}} \right)$, $\cos \left( 2{{\theta }_{1}} \right)=\left( \frac{1}{2}\frac{{{\gamma }_{2}}}{{{\gamma }_{1}}}\frac{t_{B}^{2}}{{{r}_{B}}}-\frac{1}{2}\frac{t_{B}^{2}}{{{r}_{B}}}-{{r}_{B}} \right)$ and $\sin \left( 2{{\theta }_{1}} \right)=P\frac{{{t}_{B}}\sqrt{4{{\gamma }_{1}}{{\gamma }_{2}}-t_{B}^{2}{{({{\gamma }_{1}}+{{\gamma }_{2}})}^{2}}}}{2{{\gamma }_{1}}{{r}_{B}}}$, where the factor $P=\pm 1$ represents the phase change caused by the nanocavity, $P$(=+/- 1) is the parity of the Fano resonance \cite{FANJOSAB}, and the sign can be determined by the position of the PTE (the blockade hole) relative to the PhC H0 nanocavity \cite{AOL}. In this work, the blockade hole (BH) is placed in the symmetry plane of the nanocavity, leading to a Fano resonance with a reflection spectrum of blue parity ($P=1$) because the eigen mode of the H0 nanocavity has even symmetry with respect to the mirror plane \cite{MHOL,YYLPR}. The reflection intensity and phase change of the Fano mirror are simply given as ${{R}_{R}}\left( \omega  \right)={{\left| {{r}_{R}}\left( \omega  \right) \right|}^{2}}$ and ${{\phi }_{R}}\left( \omega  \right)=\arg \left\{ {{r}_{R}}\left( \omega  \right) \right\}$. As for the left mirror, for simplicity, we assume ${{r}_{L}}\left( \omega  \right)=1$ and ${{\phi }_{L}}\left( \omega  \right)=0$.
\begin{figure}
\centering
\includegraphics[width=2.3in]{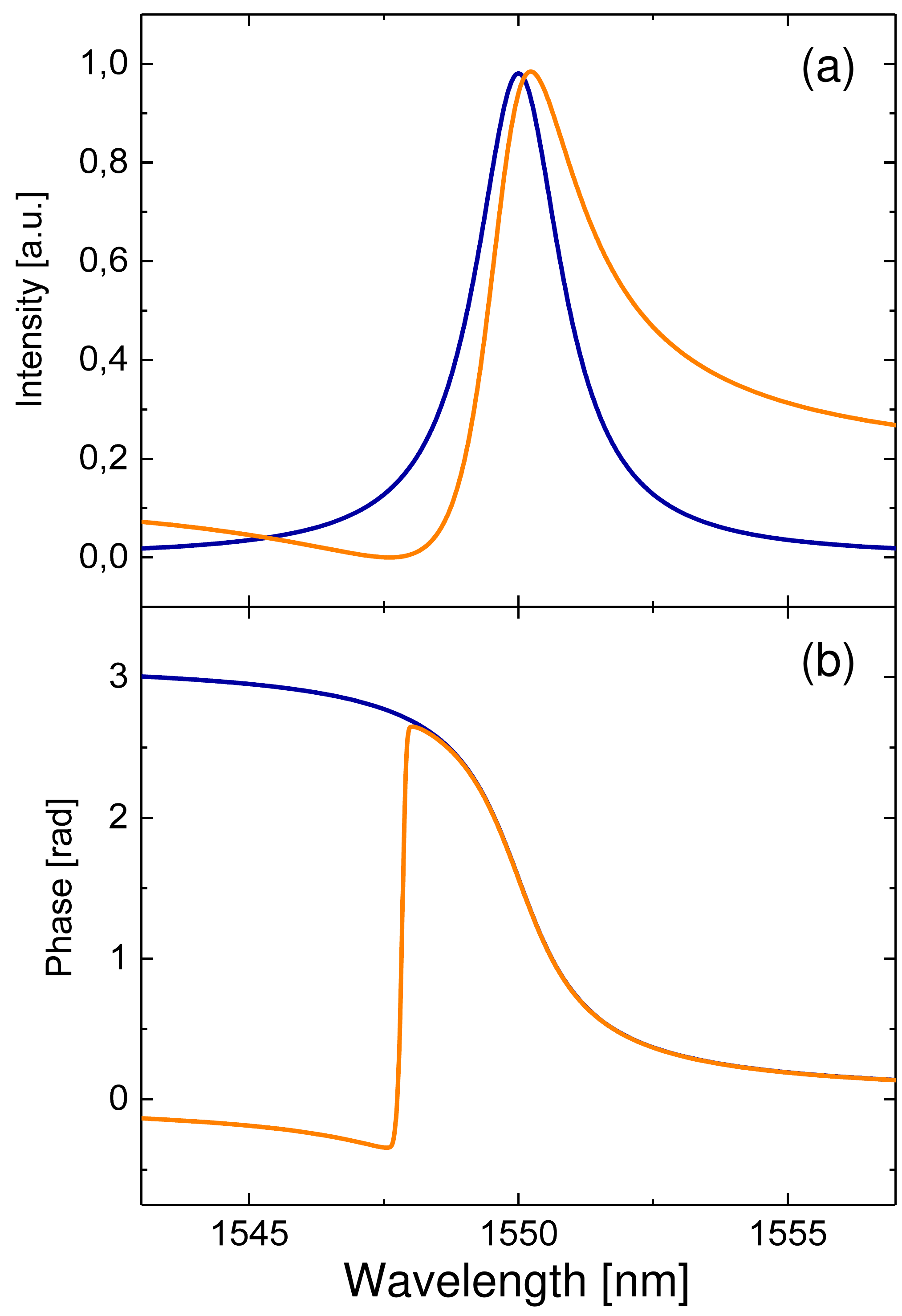}%
\caption{\label{fig:SF2} Reflection (a) intensity and (b) phase spectrum of two Fano mirrors with a symmetric Lorentzian line (${{r}_{B}}=0$, blue curve) and an asymmetric Fano line (${{r}_{B}}=0.4$, yellow curve), respectively.}
\end{figure}
Fig.\ref{fig:SF2} shows an example of the reflection intensity and phase spectrum of two Fano mirrors with the same $Q$-factors for the nanocavity (${{Q}_{v}}$=100000, ${{Q}_{\alpha }}=\infty $, ${{Q}_{c}}$=800). When blue detuned, a larger transmission contrast can be achieved for the mirror with a Fano transfer function, as compared to the mirror with a Lorentzian transfer function, for the same resonance shift, while the phase changes are almost the same.\\

\noindent \textbf{A.2. Laser dynamics}

\noindent In the L-cavity, the propagation constant $k$ is a complex wavenumber
\begin{equation}\label{E7}
\renewcommand\theequation{s\arabic{equation}}
k(\omega ,N)=\frac{\omega }{c}{{n}_{b}}\left( \omega  \right)+\Delta k\
\end{equation}
where $\Delta k$ is a perturbation introduced by the active region and enhanced by slow-light effects \cite{JMOL}. For a fixed frequency $\omega $, it depends on the carrier density $N$. Neglecting the spatial variation of carrier density along the L-cavity length, $\Delta k$ can be written by reformulating the index dependence of the carrier density via the linewidth enhancement factor $\alpha =-\frac{4\pi }{\lambda }\frac{dn/dN}{dg/dN}$ as
\begin{equation}\label{E8}
\renewcommand\theequation{s\arabic{equation}}
\Delta k=-\sigma \frac{i}{2}[\left( 1-i\alpha  \right)\Gamma {{g}_{N}}(N-{{N}_{tr}})-{{\alpha }_{i}}]\
\end{equation}
In Eqs. (s\ref{E7}) and (s\ref{E8}), ${{n}_{b}}={{n}_{0}}+\Delta {{n}_{T}}$ is the background refractive index with ${{n}_{0}}$ being the phase index of the PhC WG and $\Delta {{n}_{T}}$ being the phase index change due to the thermal effects and it can be tuned by varying slightly the pump spot position around the L-cavity. $\sigma ={{n}_{g}}/{{n}_{b}}$ is the slow-down factor, $\Gamma $ is the optical confinement factor, ${{g}_{N}}$ is the differential gain, and ${{\alpha }_{i}}$ is the internal propagation loss which depends on the slow-down factor \cite{PhCLoss_OE,JMFL}.
The boundary conditions for the complex electrical fields propagating in a laser cavity can be formulated at a reference plane \cite{TROMBORG_JQE}, which we take to be just left of the nanocavity
\begin{equation}\label{E9}
\renewcommand\theequation{s\arabic{equation}}
{{E}^{+}}\left( \omega  \right)={{r}_{1}}\left( \omega  \right){{E}^{-}}\left( \omega  \right)+{{F}_{L}}\left( \omega  \right)\
\end{equation}
\begin{equation}\label{E10}
\renewcommand\theequation{s\arabic{equation}}
{{E}^{-}}\left( \omega  \right)={{r}_{R}}\left( \omega  \right){{E}^{+}}\left( \omega  \right)\
\end{equation}
here ${{r}_{1}}\left( \omega  \right)={{r}_{L}}\left( \omega  \right)\exp[2ik(\omega ,N)L]$, ${{E}^{+}}\left( \omega  \right)$ (${{E}^{-}}\left( \omega  \right)$) is the forward (backward) propagating electrical field in the L-cavity, ${{F}_{L}}(\omega )$ is an effective Langevin noise source. Neglecting spontaneous emission noise, the steady-state lasing oscillation condition can be expressed
\begin{equation}\label{E11}
\renewcommand\theequation{s\arabic{equation}}
{{r}_{L}}(\omega ){{r}_{R}}(\omega )\exp[2ik(\omega ,N)L]=1\
\end{equation}
where $L$ is the effective L-cavity length. From Eq. (s\ref{E11}), the amplitude and phase matching condition can be obtained straightforwardly as
\begin{equation}\label{E11_2}
\renewcommand\theequation{s\arabic{equation}}
\begin{aligned}
  & \Gamma g(N)=\Gamma {{g}_{N}}(N-{{N}_{tr}}) \\
 & ={{\alpha }_{i}}+\frac{1}{2\sigma L}\ln \left( \frac{1}{{{\left| {{r}_{L}}(\omega ) \right|}^{2}}{{\left| {{r}_{R}}(\omega ) \right|}^{2}}} \right) \\
\end{aligned}\
\end{equation}
\begin{equation}\label{E12}
\renewcommand\theequation{s\arabic{equation}}
\arg \left\{ {{r}_{L}}(\omega ) \right\}+\arg \left\{ {{r}_{R}}(\omega ) \right\}+\frac{2\omega }{c}n\left( \omega ,N \right)L=2m\pi \
\end{equation}
here $m$ is an integer and $n\left( \omega ,N \right)=-c\sigma \alpha \Gamma {{g}_{N}}(N-{{N}_{tr}})/2\omega +{{n}_{b}}\left( \omega  \right)$. Neglecting the nonlinearity in the nanocavity, the steady state solution $(\omega ,N)=({{\omega }_{s}},{{N}_{s}})$ can be obtained by solving Eqs. (s\ref{E6}), (s\ref{E11_2}) and (s\ref{E12}) numerically. Noting that more than one solution can be obtained, however, the dominant lasing mode will be the solution with the lowest carrier density.

Next, to investigate the laser dynamics, we employ Taylor expansion to the first order around the steady state point so that
\begin{equation}\label{E13}
\renewcommand\theequation{s\arabic{equation}}
\begin{aligned}
  & k\left( \omega ,N \right)\simeq k\left( {{\omega }_{s}},{{N}_{s}} \right)+{{\left. \frac{\partial k\left( \omega ,N \right)}{\partial \omega } \right|}_{{{\omega }_{s}},{{N}_{s}}}}\left( \omega -{{\omega }_{s}} \right) \\
 & \ \ \ \ \ \ \ \ \ \ \ \ \ \ +{{\left. \frac{\partial k\left( \omega ,N \right)}{\partial N} \right|}_{{{\omega }_{s}},{{N}_{s}}}}\left( N-{{N}_{s}} \right) \\
\end{aligned}\
\end{equation}
Inserting Eq. (s\ref{E13}) into Eq. (s\ref{E9}) and taking the Fourier transformation on both sides of Eqs. (s\ref{E9}) and (s\ref{E10}), i.e. ${{A}^{\pm }}(t){{e}^{-j{{\omega }_{s}}t}}=\frac{1}{\sqrt{2\pi }}\int_{-\infty }^{\infty }{{{E}^{\pm }}(\omega ){{e}^{-j\omega t}}d\omega }$, we finally arrive at a differential equation governing the lasing mode as
\begin{equation}\label{E14}
\renewcommand\theequation{s\arabic{equation}}
\begin{aligned}
  & \frac{d{{A}^{+}}(t)}{dt}=\left( \frac{1}{2}\left( 1-j\alpha  \right)\sigma {{G}_{N}}\Delta N\left( t \right)-\frac{1}{{{\tau }_{in}}} \right){{A}^{+}}(t) \\
 & \ \ \ \ \ \ \ \ \ \ \ \ \ +\frac{1}{{{r}_{2}}\left( {{\omega }_{s}},{{N}_{s}} \right){{\tau }_{in}}}{{A}^{-}}(t) \\
\end{aligned}\
\end{equation}
Here we have neglected the noise term ${{F}_{L}}(\omega )$, as well as the dispersion of the slow-down factor, the material gain and the left mirror, which are all reasonable approximations for the structure investigated experimentally. In Eq. (s\ref{E14}), ${{\tau }_{in}}=2L/{{\upsilon }_{g}}$ is the cavity round-trip time with ${{\upsilon }_{g}}=c/{{n}_{g}}$ being the group velocity of the PhC WG, $\Delta N(t)$ is the deviation of the carrier density from the steady state ${{N}_{s}}$, and ${{G}_{N}}=\Gamma {{\upsilon }_{g}}{{g}_{N}}$. The power of the input/output field to/from the Fano mirror is ${{P}^{\pm }}(t)={{\left| \sqrt{2{{\varepsilon }_{0}}nc}{{A}^{\pm }}(t) \right|}^{2}}={{\left| s_{1}^{\pm }(t) \right|}^{2}}$. In addition, by multiplying ${{e}^{j{{\theta }_{1}}}}$ on both sides of Eq. (s\ref{E1}), defining ${{A}_{c}}(t)={{e}^{j{{\theta }_{1}}}}a(t)/\sqrt{2{{\varepsilon }_{0}}nc}$ and replacing $s_{1}^{\pm }(t)$ with $\sqrt{2{{\varepsilon }_{0}}nc}{{A}^{\pm }}(t)$, Eqs. (s\ref{E1})-(s\ref{E3}) can be transformed as

\begin{equation}\label{E16}
\renewcommand\theequation{s\arabic{equation}}
\frac{d{{A}_{c}}(t)}{dt}=(-i{{\delta }_{c}}-{{\gamma }_{T}}){{A}_{c}}(t)+\sqrt{2{{\gamma }_{1}}}{{e}^{i2{{\theta }_{1}}}}{{A}^{+}}(t)\
\end{equation}
\begin{equation}\label{E17}
\renewcommand\theequation{s\arabic{equation}}
{{A}^{-}}(t)={{r}_{B}}{{A}^{+}}(t)+\sqrt{2{{\gamma }_{1}}}{{A}_{c}}(t)\
\end{equation}
\begin{equation}\label{E18}
\renewcommand\theequation{s\arabic{equation}}
s_{2}^{-}(t)/\sqrt{2{{\varepsilon }_{0}}nc}=-i{{t}_{B}}{{A}^{+}}(t)+\sqrt{2{{\gamma }_{2}}}{{e}^{i{{\theta }_{2}}-i{{\theta }_{1}}}}{{A}_{c}}(t)\
\end{equation}
Furthermore, the carrier dynamics in the laser L-cavity is governed by the traditional rate equation as
\begin{equation}\label{E19}
\renewcommand\theequation{s\arabic{equation}}
\frac{dN(t)}{dt}=\frac{\eta {{P}_{p}}(t)}{\hbar {{\omega }_{p}}{{V}_{p}}}-\frac{N(t)}{{{\tau }_{c}}}-{{\upsilon }_{g}}\Gamma g(N(t))\frac{I(t)}{{{V}_{C}}}\
\end{equation}
where $N(t)={{N}_{s}}+\Delta N(t)$, ${{P}_{p}}(t)$ is the power of the pump light with a frequency of ${{\omega }_{p}}$, $\eta $ is the pump efficiency, and ${{V}_{p}}$ is the pump volume (pump spot size), ${{V}_{C}}=\Gamma AL$ is the volume of the active region with $A$ being the cross sectional area of the WG mode. $I(t)$ is the photon number within the L-cavity which can be estimated according to the optical field ${{A}^{+}}(t)$ as $I(t)={{\sigma }_{s}}({{\omega }_{s}},{{N}_{s}}){{\left| {{A}^{+}}(t) \right|}^{2}}$ with ${{\sigma }_{s}}({{\omega }_{s}},{{N}_{s}})$ being a parameter that can be calculated from the steady state solution as \cite{TROMBORG_JQE}
\begin{equation}\label{E20}
\renewcommand\theequation{s\arabic{equation}}
\begin{aligned}
  & {{\sigma }_{s}}({{\omega }_{s}},{{N}_{s}})= \\
 & \frac{2{{\varepsilon }_{0}}n{{n}_{g}}}{\hbar {{\omega }_{s}}}[\frac{\left( \left| {{r}_{L}}\left( {{\omega }_{s}} \right) \right|+\left| {{r}_{R}}\left( {{\omega }_{s}} \right) \right| \right)\left( 1-\left| {{r}_{L}}\left( {{\omega }_{s}} \right) \right|\left| {{r}_{R}}\left( {{\omega }_{s}} \right) \right| \right)}{\sigma \left( \Gamma g\left( {{\omega }_{s}},{{N}_{s}} \right)-{{\alpha }_{i}} \right)\left| {{r}_{L}}\left( {{\omega }_{s}} \right) \right|} \\
 & +\frac{c}{{{\omega }_{s}}n}\frac{\left| {{r}_{L}}\left( {{\omega }_{s}} \right) \right|}{\left| {{r}_{R}}\left( {{\omega }_{s}} \right) \right|}\operatorname{Im}\left\{ {{r}_{R}}\left( {{\omega }_{s}} \right) \right\}] \\
\end{aligned}\
\end{equation}
Eq. (s\ref{E14}) derived by Taylor expansion is valid if the lasing state does not change too much from the expansion point. Carrier diffusion between the L-cavity and nanocavity is not taken into account considering the quantum confinement of QDs. Thermal dynamics is not included because of the much longer time constants involved compared to the carrier dynamics and the resulting pulsation frequencies. The parameters used for the simulations are: ${{\gamma }_{1}}={{\gamma }_{2}}$, ${{r}_{B}}=0$, ${{Q}_{c}}=780$, ${{Q}_{\alpha }}=1134$, ${{Q}_{v}}=100000$, $L=5.48\upmu$m, $n$=3.5, $\sigma=3$, $m=25$,  $A=0.105\upmu$m$^{2}$, $V_{p}=0.576\upmu$m$^{3}$, ${{g}_{N}}=3.8\times {{10}^{-19}}$m$^{2}$, ${{N}_{tr}}=2\times {{10}^{23}}$m$^{-3}$, ${{\tau }_{c}}=0.28$ns, ${{\alpha }_{i}}=1000$m$^{-1}$, $\alpha=0.2$, $\Gamma =0.163$. The nonlinear coefficients in the nanocavity are ${{K}_{D}}={{\Gamma }_{c}}{{{\upsilon }'}_{g}}\alpha {{g}_{N}}/2$, ${{K}_{A}}=-{{\Gamma }_{c}}{{{\upsilon }'}_{g}}{{g}_{N}}/2$, ${{G}_{LA}}={{g}_{N}}{{{\upsilon }'}_{g}}/\left( \hbar \omega {{V}_{pc}} \right)$ with ${{{\upsilon }'}_{g}}=c/6{n}$ being an effective group velocity of the nanocavity, ${{\Gamma }_{c}}=0.98$ being the confinement factor of the nanocavity and ${{V}_{pc}}=0.22\upmu$m$^{3}$ being an effective nanocavity mode volume. These parameters are chosen based on ordinary laser material parameters and FDTD/finite element method calculations.
\begin{figure*}
\centering
\includegraphics[width=6in]{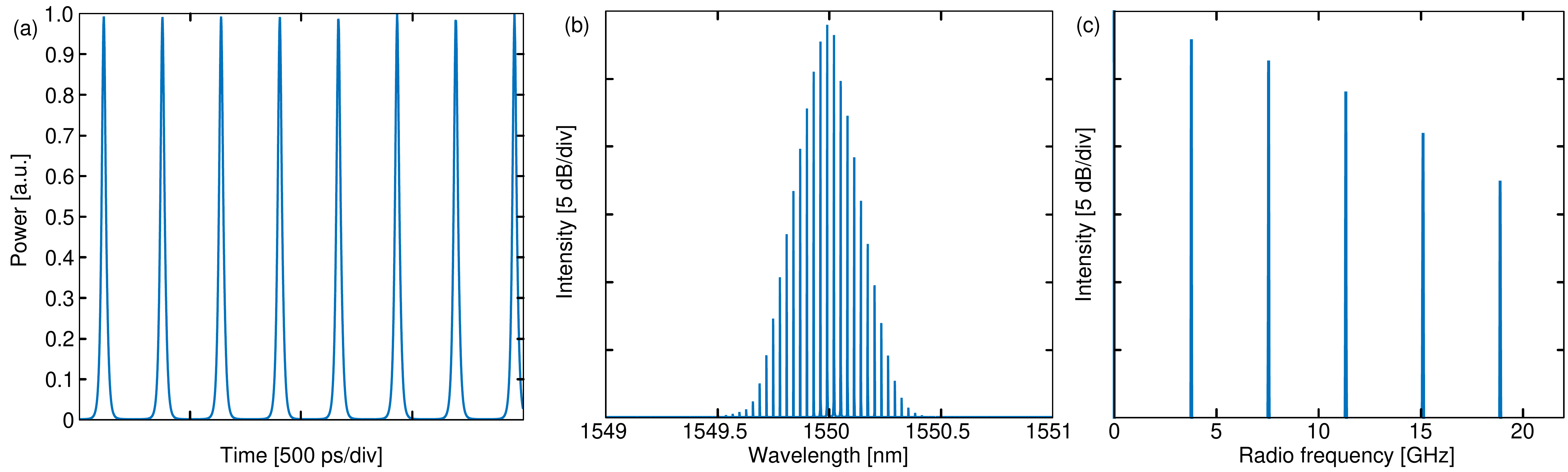}%
\caption{\label{fig:SF3} Simulated (a) optical pulses $|s_{2}^{-}(t){{|}^{2}}$ propagating in the output PhC WG and corresponding (b) optical and (c) radio frequency spectrum. Here ${{\omega }_{s}}={{\omega }_{0}}$ and the pump power is set as 1.35$P_{0}$ where $P_{0}$ is the lowest pumping power where self-pulsing sets in.}
\end{figure*}
\begin{figure*}
\centering
\includegraphics[width=4.5in]{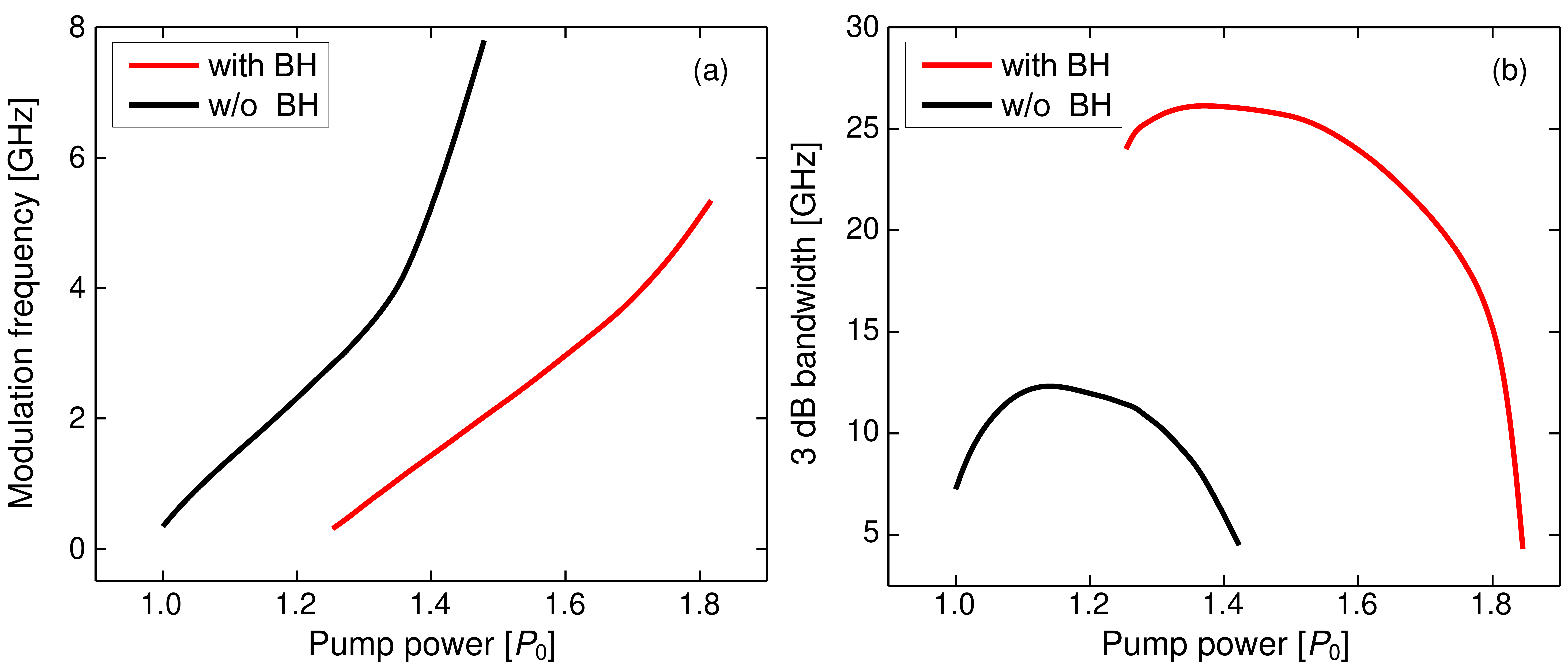}%
\caption{\label{fig:SF4} Simulated (a) modulation frequency and (b) 3dB bandwidth of optical spectra versus pump power. Red and black lines correspond to a FL with ($r_{B}$=0.4) and without ($r_{B}$=0) a blockade hole in the WG, respectively.}
\end{figure*}

Figs. \ref{fig:SF3} and \ref{fig:SF4} show examples of the calculated lasing spectrum and output pulses in the self-pulsation regime. The dependence of the the modulation frequency and 3 dB optical bandwidth agree qualitatively with the experimental observations reported in Fig. 4. In both experiment and simulations (Fig. \ref{fig:SF3}), the optical spectra always show single-mode characteristic, i.e., a single central peak with modulation sidebands that grow with power in the pulsation regime, clearly distinguishing the laser dynamics from beating-type oscillations where two main modes and a single radio frequency sideband exist \cite{BEATING}. From the simulations, short pulse generations on the order of 15 ps can be expected. We found that the power range for self-pulsation to be somewhat smaller in the simulations compared to the measurements, which may be due to the neglect of other types of nonlinear effects in the nanocavity as well as the details of the temporal and spatial variation of the carrier distributions.
\begin{figure*}
\centering
\includegraphics[width=6.5in]{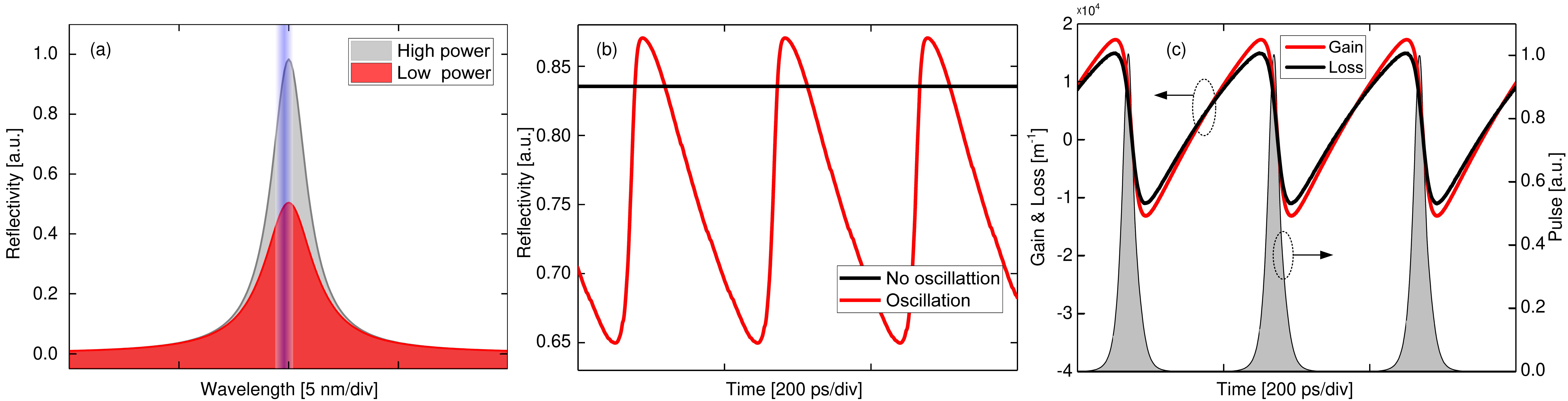}%
\caption{\label{fig:SF5} (a) Schematic view of reflection spectrum of the Fano mirror when the lasing power is low (red) and high (gray). The blue shaded region indicates the frequency where the laser self-pulses. (b) An example of the simulated time evolution of the mirror reflectivity. The black (red) curve corresponds to the case before (after) self-pulsations sets in. (b) An example of the simulated time evolution of the gain (red curve), loss (black curve) changes and laser output (gray shade) of the FL in the regime of self-pulsations.}
\end{figure*}

To understand the physics behind the self-pulsation, we model the time evolution of the mirror reflectivity of the Fano mirror, as well as the gain and loss change of the system, as shown in Fig. \ref{fig:SF5}. From Eq. (s\ref{E14}), the gain and loss variation can be defined as $\frac{1}{2}\sigma {{G}_{N}}\Delta N\left( t \right)/{{\upsilon }_{g}}$ and $\operatorname{Re}\left\{ \frac{1}{{{\tau }_{in}}{{\upsilon }_{g}}}-\frac{1}{{{r}_{2}}\left( {{\omega }_{s}},{{N}_{s}} \right){{\tau }_{in}}{{\upsilon }_{g}}}\frac{{{A}^{-}}(t)}{{{A}^{+}}(t)} \right\}$, respectively. We find that the $Q$-factor change of the nanocavity plays a key role (Fig. \ref{fig:SF5}(a)). If the power in the nanocavity experiences a transient increase, due to spontaneous emission in the laser cavity, the QDs in the nanocavity are saturated further, and the linear absorption and the mirror loss are decreased, leading to an increase of the mirror reflectivity, cf. Figs. \ref{fig:SF5}(b) and \ref{fig:SF5}(c). This provides positive feedback to the optical power, which grows further, and the system settles instead in a dynamical state where the gain and loss vary in time (Fig. \ref{fig:SF5}(c)), resulting in the generation of a train of short optical pulses. The nanocavity thus functions as a saturable absorber and the pulse generation is an example of passive $Q$-switching. Considering the low power-density of nanolasers, the use of a nanocavity to realize the Fano resonance is key to the pulse generation: It is the highly localized cavity field, enabled by the large $Q/V$-ratio, that enables the achievement of a saturable mirror even for low power levels. In the simulations, self-pulsations occur when the lasing wavelength is in the vicinity of the nanocavity resonance peak (see the blue region in Fig. \ref{fig:SF5}(a)) and this area is a bit blue-shifted with respect to the resonance wavelength due to the free carrier nonlinearity.

\section{B.	Experiments}
\noindent \textbf{B.1. Photonic-crystal sample and design}

\noindent The samples are fabricated on a 250 nm thick InP membrane which contains three layers of InAs quantum dots with effective thickness of 1.65 monolayers, separated by 30 nm thick InP. Photoluminescence spectra of the InP wafer shows a spectrum centred at 1558 nm with a full width at half maximum of $\sim$166 nm. The InP structure is bonded to a silicon carrier using Bisbenzocyclobutene with a 1 $\upmu$m silicon dioxide layer. Details about the fabrication process can be found in \cite{WQ_OE}. The photonic-crystal structure has a hexagonal lattice of holes with lattice constant $a$=455 nm and hole radius $R$=110 nm. A line-defect cavity is formed by side coupling an H0-type nanocavity with a standard W1-type (defined by removing a single row of air holes) waveguide, albeit with the innermost array of holes being shifted towards the waveguide center by 0.1$a$ to enhance the coupling to the nanocavity, leading to a slow-down factor of $\sim$3 around the operation wavelength. The H0 nanocavity is formed by shifting two neighboring air holes 0.16$a$ in opposite directions horizontally (parallel to the waveguide) and two other neighboring air holes 0.15$a$ in opposite directions vertically (perpendicular to the waveguide), leading to a coupling $Q$-factor of $\sim$800 and a vertical $Q$-factor of $\sim$100000. To tune the transmittance of the continuum mode, a blockade hole with a radius of $R_{B}=0.8R$ is added in the waveguide below the nanocavity. This gives a reflection coefficient of $r_{B}^{2}=0.16$ around the H0 nanocavity resonance, as estimated by FDTD simulations. For the broad-band left mirror, one air hole is shifted towards the left by 0.21$a$ to reduce the scattering loss of the left mirror \cite{NODA_Q}. The structure is not cleaved and the right end of the photonic-crystal waveguide is directly terminated in the non-structured region (interface reflection is $\sim$0.02). \\

\noindent \textbf{B.2. Experimental set-up}

\noindent The laser samples are vertically pumped with a 1480 nm laser diode using a micro-photoluminescence set-up \cite{WQ_OE}, with precise control of pump position and area, and monitored by an infra-red camera (Xeva-1.7-320). The full width at half maximum (diameter) of the pump spot is fixed at $\sim$3 $\upmu$m. The emission from the photonic-crystal sample is collected vertically using the same objective lens, with a numerical aperture of 0.65. The transmission efficiency from the pump diode to the objective is $\sim$30\% and the vertical collection efficiency is $\sim$20\% (estimated by comparing the farfield pattern of the laser emission with the lens¡¯s numerical aperture area), which may vary slightly for different cavity structures. All measurements are performed with continuous-wave injection at room temperature. For the output spectrum measurements, after being isolated from the reflected pump beam by a long-pass filter, the signal is analyzed using an optical spectrum analyzer (YOKOGAWA AQ6370D), or a highly sensitive spectrometer (Princeton Instruments, Acton SP2500) with a cooled low-noise InGaAs one-dimensional detector array. For the radio frequency measurements, the collected output signal is first amplified using a low-power erbium doped fiber amplifier (Amonics AEDFA-L-PA-35-B-FA) cascaded with a tunable optical band-pass filter (Santec OTF-350) followed by an optical isolator, after which the signal is reamplified using a standard in-line erbium doped fiber amplifier, filtered using another tunable optical band-pass filter, and detected with a photodiode and monitored with a radio frequency spectrum analyzer (HP 70000 SYSTEM) having a bandwidth of 22 GHz. For the optical and radio frequency spectrum measurements, the spectrum scan speed was chosen as a good compromise between the need to obtain a sufficiently high signal-to-noise ratio, and the requirement to minimize thermal drift.

\noindent \textbf{B.3. Lasing mode identification}
\begin{figure*}
\centering
\includegraphics[width=4.8in]{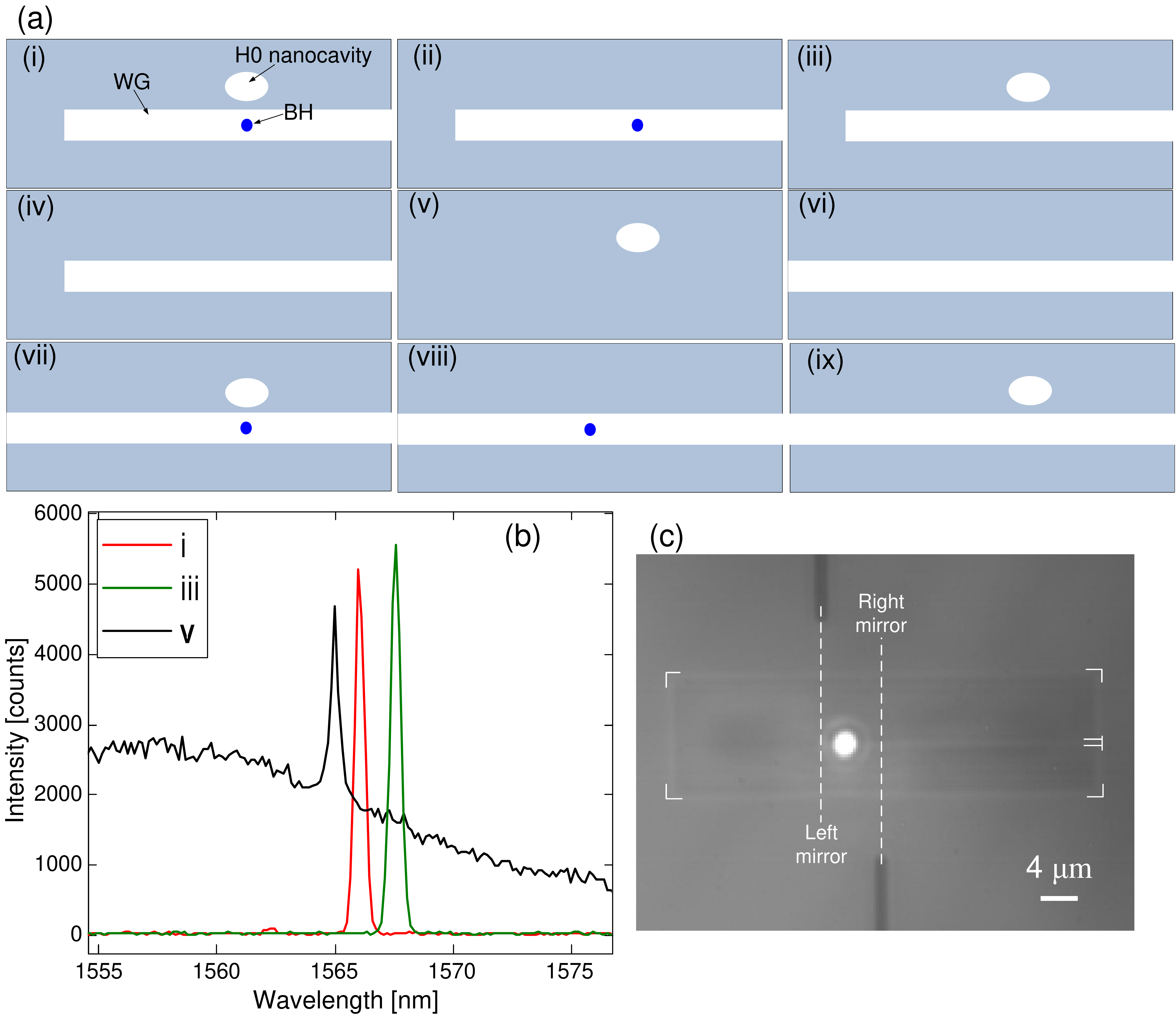}%
\caption{\label{fig:SF6} (a) Schematic of nine different PhC cavity-waveguide structures. (b) Lasing spectra of a FL with a blockade hole (BH) (sample (i), red curve), a FL without BH (sample (iii), green curve) and a corresponding isolated H0 nanocavity (sample (v), black curve). The spectrometer integration time was two orders of magnitude larger for the black curve as compared to the red and green curves. (c) Infrared camera image taken when the FL is pumped by a continuous wave light source (white dot) within the line-defect cavity region. Two trenches (black rectangles) are patterned near the FL to indicate the left and right mirror positions (white dashed lines), respectively. The white solid lines highlight the FL boundary and the waveguide end position.}
\end{figure*}

\noindent To verify that the lasing mode originates from the Fano interference, we fabricated various PhC samples with the same structural parameters and on the same wafer (and processed in the same batch), as schematically shown in Fig.\ref{fig:SF6}(a). Only the structures (i), (iii) and (v) were able to lase with the pump power available. Sample (i) and (iii) are pumped within the L-cavity region between the left mirror and the H0 nanocavity, cf. Fig.\ref{fig:SF6}(c), confirming that the FL eigenmode (sample (i) and (iii)) relies on the Fano interference between the PhC WG and the nanocavity. Note, in particular, that the structures (vii) and (ix) do not undergo a transition to lasing: The coupling between the nanocavity and the waveguide reduces the photon lifetime so much that the gain of the QDs cannot compensate the cavity loss.

\subsection{}
\subsubsection{}
\bibliographystyle{ieeetr}
\bibliography{FLBIB}

\end{document}